\documentclass{ws-procs975x65}

\begin{document}

\title{WHY WE NEED DARK ENERGY}
\author{D. PAV\'{O}N$^*$}
\address{Departamento de F\'{\i}sica, Universidad Aut\'{o}noma de Barcelona,\\
Bellaterra, 08193, Spain\\
$^*$E-mail: diego.pavon@uab.es}

\author{N. Radicella$^*$}
\address{Dipartimento di Fisica ``E.R. Caianiello", Universit\`{a} di Salerno,
I-84084 Fisciano, Italy}
\begin{abstract}
It is argued that dark energy -or something dynamically equivalent
at the background level- is necessary if the expanding universe is
to behave as an ordinary macroscopic system; that is, if it is to
tend to some thermodynamic equilibrium state in the long run.
\end{abstract}

\keywords{Cosmology; Dark energy; Thermodynamics}

\bodymatter

\section{Introduction}
\label{sec:intro}
The standard cosmological cold dark matter model  was in good
health until around the last decade of the previous century when
it was realized that the fractional density of matter falls well
below unity. Its full dismissal came shortly after, about the
close of the century, with the discovery of the current cosmic
acceleration. However, to account for the acceleration in
homogeneous and isotropic models one must either introduce some
exotic energy component with a huge negative pressure (called dark
energy) or, more radically, devise some theory of gravity more
general than Einstein's. Thus, both solutions look somewhat forced
and not very aesthetical. Here we very briefly argue that dark
energy (or something equivalent) is demanded on thermodynamic
grounds. That is to say, we provide what we believe is a sound
thermodynamic motivation for the existence of dark energy. Details
can be found in Ref.~\refcite{grg_nd1}.

\section{Simple models}
\label{sec:simple}
Isolated, ordinary, macroscopic systems spontaneously tend to
thermodynamic equilibrium. This constitutes the empirical basis of
the second law of thermodynamics. Succinctly, the latter asserts
that the entropy, $S$,  of such systems can never decrease, $S'
\geq 0$, and  that it must be concave, $S'' < 0$, at least during
the last stage of approaching equilibrium \cite{callen}. It seems
worthwhile to explore the consequences of this law when applied to
spatially flat Friedmann-Robertson-Walker universes.

The entropy is contributed by two terms, the entropy of the
apparent horizon, $S_{A} = k_{B} {\cal A}/(4 \ell_{Pl}^{2})= k_{B}
\pi/H^{2}$ (where ${\cal A}$ denotes the horizon area), and the
entropy of matter and fields, $S_{f}$, enclosed by it. Let $\rho$
and $w$ be the energy density and equation of state parameter
(assumed constant) of the latter. Then, $S'_{A} \propto (1+w)/(a
\rho)$ and $S''_{A} \propto (1+w)(2+3w)/(a^{2} \rho)$ (the prime
means derivative with respect to the scale factor). Likewise,
$S'_{f} \propto (1+w)(1+3w)/(aH)$ and $S''_{f} =
(1+w)(1+3w)(1+w)a^{(9w-1)/2}$.  These expressions reveal that
radiation dominated universes cannot tend to thermodynamic
equilibrium in the long run. The same is true for universes
dominated either by matter or phantom energy. However, for dark
energy dominated universes with $-1 \leq w < -2/3$, one has
$S'_{A} + S'_{f} > 0$ at all times, and $S''_{A} + S''_{f} < 0$ at
least when $a \rightarrow \infty$. This suggests the necessity of
non-phantom dark energy, or some modified gravity scenario able to
lead to a suitable acceleration at late times, for the Universe
to behave as an ordinary macroscopic system.

\section{More general models}
\label{sec:moregm}
Thus far we have restricted ourselves to cosmological models with
$w =$ constant and governed by Einstein gravity. This section very
briefly reviews a handful of more general models that show overall
compatibility with the observational constraints.

$(i)$ For the model of Barboza and Alcaniz \cite{alcaniz}, in
which the universe is dominated by pressureless matter and dark
energy with equation of state $w(z) = w_{0} + w_{1}
z(1+z)(1+z^{2})^{-1}$, the second law is fulfilled either if
$w_{1} < 2/3$ and $-1< w_{0} < -2/3$, or if $2/3 \leq w_{1} < 1$
and $-1 < w_{0}< -w_{1}$.

$(ii)$ In the original Chaplygin gas model \cite{chaplygin} ($p =
-A/\rho$, $\rho = \sqrt{A +(B/a^{6})}$), that unifies matter and
dark energy, $S'_{f} <0$ and $S''_{f} >0$ when $a \rightarrow
\infty$; nevertheless, $S' > 0 $ and $S'' < 0$ in the same limit.

$(iii)$ The holographic dark energy model of
Ref.~\refcite{holographic} in which the dark energy density varies
as the area of the apparent horizon and decays into matter in such
a way that the ratio of both densities is a constant (thus solving
the coincidence problem), is also  seen to tend to thermodynamic
equilibrium at late times.

$(iv)$ In the modified gravity model of Dvali {\em et al.}
\cite{dgp} our four dimensional Universe is considered a brane
embedded in a five dimensional bulk with flat Minkowski metric.
There is no dark energy, just matter and radiation. The condition
for having $S' >0 $ and $S'' <0$ as $a \rightarrow \infty$ reduces
to a bound on the current number of dust particles, which is
fulfilled by a huge margin.

$(v)$ The Cardassian model, proposed by Freese and Lewis
\cite{katherine}, also dispenses altogether with dark energy but
introduces an extra term, proportional to $\rho^{\alpha}$, in the
usual Friedamnn's equation. Because it can be mapped to a dark
energy model with $w = \alpha -1$, it presents a healthy
thermodynamic behavior provided $\alpha$ lies in the range $(0,
2/3)$.

$(vi)$ Lastly, torsion gravity replaces the scalar curvature in
Einstein relativity by the torsion $\tau$, the action being
\[
I = \frac{1}{16\pi G}\int{d^{4}x \, \sqrt{-g} (\tau \, +\,
f(\tau))} \, + \, I_{matter}\, .
\]
In the  of Bengochea and Ferraro \cite{ben} $f(\tau) = -
\alpha(-\tau)^{-n}$. Using the best fit values for $\alpha$ and
$n$, it is found that $S' <0$ and that $S'' >0$ in long run,
thereby the universe does not tend to thermodynamic equilibrium.

\section{Conclusions}
Altogether, $(i)$ neither a radiation dominated nor a cold dark
matter dominated universe can tend to thermodynamic equilibrium in
the long run. However, dark energy dominated universes with
constant $w$ in the range $-1 \leq w <-2/3$ and the $\Lambda$CDM
model can. $(ii)$ Phantom models with $w =$ constant, cannot.
$(iii)$ Dark energy models with $w \neq$ constant deserve a
separate analysis. In particular, the Chaplygin gas model
\cite{chaplygin} can, as well as the model of Barboza and Alcaniz
\cite{alcaniz}  for an substantial range of its parameters, and
some holographic models \cite{holographic}. $(iv)$ Several
modified gravity models, such as DGP \cite{dgp} and Cardassian
models \cite{katherine}, can, but the model of Bengochea and
Ferraro \cite{ben} cannot. $(v)$ Models in which the present
accelerated stage of expansion is just transitory conflict with
the second law. $(vi)$ The entropy of the Universe seems to tend
to some maximum value (of about $H^{-2}$ when $a \rightarrow
\infty$), but in order to reach a firmer conclusion, accurate
measurements regarding $H(z)$ are called for. $(vii)$ Finally, the
existence of dark energy or -equivalently- some modified gravity
theory could have been expected on thermodynamic grounds.

\section*{Acknowledgments}
This research was partially supported by the Spanish Ministry of
Science and Innovation under Grant FIS2009-13370-C02-01, and the
``Direcci\'{o} de Recerca de la Generalitat" under Grant
2009SGR-00164.


\end{document}